\documentclass[11p,a4paper]{article}
\pdfoutput=1
\usepackage{jheppub}
\usepackage[utf8]{inputenc}
\usepackage{latexsym,amsfonts,amsmath,amscd,amssymb,epsf}
\usepackage{hyperref}
\usepackage{soul,xcolor}
\usepackage{natbib}
\usepackage{graphicx}

\newcommand{\beq}{\begin{equation}}
\newcommand{\eeq}{\end{equation}}
\newcommand{\cK}{{\cal K}}
\newcommand{\cP}{{\cal P}}

\renewcommand\Re{\operatorname{Re}}

\usepackage{bm}
\let\vec=\bm
\title{From soft to hard radiation: the role of  multiple scatterings in medium-induced gluon emissions}

\author[a]{Carlota Andres,}
\author[b]{Fabio Dominguez,}
\author[b]{and Marcos Gonzalez Martinez}
\emailAdd{carlota@lip.pt}
\emailAdd{fabio.dominguez@usc.es}
\emailAdd{marcosg.martinez@usc.es}

\affiliation[a]{LIP, Av. Prof. Gama Pinto, 2, P-1649-003 Lisboa, Portugal}
\affiliation[b]{Instituto Galego de F\'isica de Altas Enerx\'ias IGFAE, Universidade de Santiago de Compostela, E-15782 Santiago de Compostela (Galicia-Spain)}
\date{January 2020}

\begin{document}

\abstract{
A proper understanding of the physics of medium-induced gluon emissions is known to be of critical importance to describe the properties of strongly interacting matter under extreme conditions. In this regard, many theoretical efforts have been directed towards obtaining analytical calculations which might help us discerning the underlying physical picture and the dominant dynamics for different regimes. These analytical approaches rely on approximations whose validity is analyzed here by comparing their results with a recently developed numerical evaluation which includes all-order resummation of multiple scatterings. More specifically, by quantitatively comparing the energy spectrum and rates, we observe that three different regimes --- each with its corresponding physical picture --- emerge naturally from the equations: the high-energy regime where the emission process is dominated by a single hard scattering, the intermediate-energy regime where coherence effects among multiple scatterings become fundamental, and the low-energy regime where the dynamics is again dominated by a single scattering but where one must include the suppression factor due to the probability of not having any further scatterings (which is obtained through the resummation of virtual terms).}

\maketitle

\section{Introduction}

The impressive results from several years of the heavy-ion programs at the Relativistic Heavy Ion Collider (RHIC) and the Large Hadron Collider (LHC) \cite{Connors:2017ptx,Armesto:2015ioy,Roland:2014jsa,Muller:2012zq,Muller:2006ee,Frawley:2008kk,Back:2004je} combined with the possible connections to dense stellar objects \cite{Dexheimer:2020zzs} have directed a great deal of efforts into developing and perfecting fundamental theories for the description of strongly interacting matter under extreme conditions. In particular, the proper understanding of the interactions of highly energetic colored particles with QCD matter at high temperatures and high densities plays a fundamental role in extracting the properties of such media, thus giving us a window into a wider comprehension of the nature of strong interactions in multi-particle systems. Under these conditions, high-energy partons lose energy through medium-induced gluon radiation, a process which is mainly responsible for the suppression of high-energy hadrons and jets observed in heavy-ion collisions (see refs.~\cite{Armesto:2011ht,Blaizot:2015lma,Qin:2015srf} for recent theoretical reviews on jet quenching).

From a purely theoretical perspective, a complete description of medium-induced radiation poses technical challenges, given the important role of multiple coherent scatterings which give rise to the well known LPM effect. These multiple scatterings can be formally resummed under the BDMPS-Z framework \cite{Baier:1996kr, Baier:1996sk,Zakharov:1996fv,Zakharov:1997uu}, but further evaluation of observables have typically required restrictive approximations which might miss some of the relevant physics of the emission process.

In order to provide adequate tools for phenomenological analyses, two main approaches have been adopted. Either developing Monte Carlo generators which capture the main physical effects, or numerically evaluating analytical expressions obtained from simplified scenarios, like the single hard scattering \cite{Gyulassy:2000er} or harmonic oscillator approximations. More recently, there has been a renewed effort to get a better understanding of the in-medium emission spectrum, both through numerical and analytical developments \cite{Feal:2018sml,Andres:2020vxs,Mehtar-Tani:2019tvy,Mehtar-Tani:2019ygg,Barata:2020sav}, which is expected to help us gain insight into our interpretation of such radiation processes, and has already shown promising results \cite{Feal:2019xfl} in elucidating the origin of effects like the centrality/energy puzzle \cite{Andres:2016iys,Burke:2013yra,Andres:2017awo}  observed in the data for single-inclusive hadron suppression.

Instead of following the usual course of employing analytical approximations to develop new numerical tools for phenomenological purposes, the aim here is to provide the feedback loop by using our recently developed framework \cite{Andres:2020vxs} to check the validity of the approximations used in some of the analytical approaches. In particular, we are interested in illuminating the discussion around the accuracy of the  multiple soft scattering result and  analyzing which are the regions where a single scattering is sufficient to describe the in-medium emission spectrum.

In our previous publication \cite{Andres:2020vxs} the emphasis was on showing that our formalism is suitable to improve phenomenological tools while here we focus on getting a better understanding of the physical picture behind the radiation process. More specifically, we study the energy spectrum integrated over all transverse momenta (no kinematical constraint) in order to provide a cleaner comparison with analytical results, even though it has to be kept in mind that for realistic conditions the phase space must be restricted. In this scenario, our framework is analogous to those in \cite{Zakharov:2004vm,CaronHuot:2010bp}, but we compute both the spectrum as a function of energy and the emission rates as a function of the medium length, while in \cite{CaronHuot:2010bp} only the latter are analyzed.

The paper is organized as follows: in section~\ref{sec:spectrum} we describe our all-order in-medium radiation framework derived in \cite{Andres:2020vxs} focusing on the equations needed to compute the energy spectrum without the kinematical constraint. In section~\ref{sec:scales} we outline the different physical pictures expected to be dominant across the different energy regions involved in the emission process. We present in section~\ref{sec:asymptotic} the low- and high-energy asymptotic regimes dominated by a single scattering, while in section~\ref{sec:mult} we turn our attention to the intermediate energy region  where multiple scatterings are crucial. In this latter section we make a quantitative comparison between our all-order in-medium energy spectrum and rates and those obtained with the analytic expansion in \cite{Mehtar-Tani:2019tvy}. Finally, we summarize and conclude in section~\ref{sec:conclusions}.

\section{Emission spectrum with full resummation of multiple scatterings}
\label{sec:spectrum}

Our starting point is the soft gluon $\vec{k}$-differential spectrum off a hard parton in the BDMPS-Z framework, which reads:\footnote{Throughout, we make use of bold symbols for the two-dimensional variables and we adopt the shorthand $\int_{\vec{p}}= \int d^2\vec{p}/(2\pi)^2$ for the transverse integrals in momentum space.}
\beq
\omega\frac{dI}{d\omega d\vec{k}^2} = \frac{2\alpha_s C_R}{(2\pi)^2\omega^2} \Re \int_0^\infty dt' \, \int_0^{t'} dt \,\int_{\vec{p}\vec{q}} 
 \vec{p} \cdot \vec{q} \,\,\widetilde{\cK}(t',\vec{q};t,\vec{p})
  \cP(\infty,\vec{k};t',\vec{q})\,,
 \label{eq:bdmps}
\eeq
where $\omega$ and $\vec{k}$ are, respectively, the energy (assumed much smaller than that of the parent parton) and transverse momentum of the emitted soft gluon, and $C_R$ is the Casimir factor corresponding to the emitter.  $\widetilde\cK (t',\vec{q};t,\vec{p})$ is the emission kernel in momentum space and $\cP(\infty,\vec{k};t',\vec{q})$ is the transverse momentum broadening factor. The emission kernel satisfies the integral equation
\beq
\widetilde{\cK}(t',\vec{q};t,\vec{p}) = (2\pi)^2\delta^{(2)}(\vec{q}-\vec{p})\,e^{-i\frac{p^2}{2\omega}(t'-t)}-\frac{1}{2}\int_t^{t'} ds\,n(s)\int_{\vec{l}}\sigma(\vec{q}-\vec{l})\widetilde{\cK}(s,\vec{l};t,\vec{p})\,,
\label{eq:SDK}
\eeq
where $n(t)$ is the linear density of scattering centers and $\sigma$ the dipole cross section. The latter contains the specific details of the parton-medium interaction and can be written in momentum space in terms of the collision rate $V$ as
\beq
\sigma(\vec{q}) = - V(\vec{q}) + (2\pi)^2\delta^{(2)}(\vec{q})\int_{\vec{l}}V(\vec{l})\, .
\label{eq:sigmaV}
\eeq

In our previous paper \cite{Andres:2020vxs}, we explained in detail how to numerically evaluate this emission spectrum. We also highlighted the relevance of correctly accounting for the transverse momentum phase space, even for the case of the energy spectrum $\omega dI/d\omega$, given that for phenomenological applications it is important to implement the kinematical limit restricting the transverse momentum of the radiated gluon to be smaller than its energy. Here, for simplicity and in order to compare to analytical approaches, we focus on the case where the $\vec{k}$-integration is performed over all transverse momentum space, thus removing the momentum broadening factor --- since its integration over the whole transverse momentum plane gives $1$.\footnote{The probability of having any broadening, including none, is clearly one.}

The medium-induced energy spectrum is then
\beq
\omega\frac{dI^\mathrm{med}}{d\omega} = \frac{2\alpha_s C_R}{\omega^2} \Re \int_0^\infty dt' \, \int_0^{t'} dt \,\int_{\vec{p}\vec{q}} \, 
 \vec{p} \cdot \vec{q} \,\,\left[\widetilde{\cK}(t',\vec{q};t,\vec{p}) - \widetilde{\cK}_0(t',\vec{q};t,\vec{p}) \right] \,,
 \label{eq:enspec}
\eeq
where the vacuum contribution is explicitly subtracted through the vacuum emission kernel
\beq
\widetilde\cK_0(t',\vec{q};t,\vec{p}) = (2\pi)^2\delta^{(2)}(\vec{q}-\vec{p})\,e^{-i\frac{p^2}{2\omega}(t'-t)}\,.
\eeq
 
By plugging eq.~(\ref{eq:SDK}) into (\ref{eq:enspec}) we can perform the $t'$-integration, being left with
\beq
\omega\frac{dI^\mathrm{med}}{d\omega} = \frac{2\alpha_s C_R}{\omega} \Re\int_0^{\infty} ds\; n(s) \int_0^s dt\int_{\vec{p}\vec{q}\vec{l}}i\,\frac{\vec{p}\cdot\vec{q}}{\vec{q}^2}\sigma(\vec{q}-\vec{l})\widetilde\cK(s,\vec{l};t,\vec{p})\,.
\label{eq:enspecrinf}
\eeq
For simplicity, we will restrict the rest of our analysis to the ``brick'' case with $n(s)=n_0\Theta(L-s)$, so the upper limit of the integral over the position of the last scattering $s$ becomes $L$. The spectrum in eq.~(\ref{eq:enspecrinf}) is evaluated by considering the emission kernel as an evolution operator following the procedure described in detail in ref.~\cite{Andres:2020vxs}. At this point, three parameters are needed to describe the medium, namely $n_0$, $L$, and an additional screening mass $\mu^2$ entering in the collision rate $V$. This number of parameters can be reduced by expressing the spectrum in terms of the dimensionless variable
\beq
x=\frac{\omega}{\bar\omega_c}=\frac{2\omega}{\mu^2L}\,,
\label{eq:our_x}
\eeq
and rescaling all the dummy variables as $\vec{p}\to\vec{p}\sqrt{2\omega/L}$ for momenta, and $t\to tL$ for time. The energy spectrum is then given by
\beq
x\frac{dI^\mathrm{med}}{dx} = 4\alpha_s C_R\, \Re\int_0^1 ds\, \int_0^s dt\int_{\vec{p}}i\vec{p}\cdot\vec{F}_x(t;\vec{p})\,,
\label{eq:spec_final}
\eeq
where $\vec{F}_x$ satisfies the differential equation
\beq
\partial_t\vec{F}_x(t;\vec{p}) = -ip^2\vec{F}_x(t;\vec{p})-\frac{1}{2}n_0L\int_{\vec{p}'}\tilde\sigma(\vec{p}'-\vec{p};x)\vec{F}_x(t;\vec{p}')\,,
\label{eq:diffeqF}
\eeq
with initial condition
\beq
\vec{F}_x(0;\vec{p}) = n_0L\int_{\vec{q}}\frac{\vec{q}}{\vec{q}^2}\tilde\sigma(\vec{q}-\vec{p};x)\,,
\eeq
and $\tilde\sigma(\vec{q};x)$ is the rescaled dipole cross section $\tilde\sigma(\vec{q};x)=x\mu^2\sigma(\vec{q}\sqrt{x\mu^2})$.

It is clear from the set of equations above that the only relevant parameter for the energy spectrum when expressed in terms of $x$ is $n_0L$.

\section{Energy scales and separation of regimes}
\label{sec:scales}
For a medium with mean free path $\lambda$ and length $L$, one can expect a given emitted gluon to encounter a number of scatterings of the order of $L/\lambda$. Each scattering center can possibly transfer transverse momentum to the gluon of the order of $\tilde\mu$, with larger momentum transfers considered rare events. Gluons which have not had any rare hard scatterings are then expected to have a transverse momentum in the range
\beq
\tilde \mu^2\lesssim \vec{k}^2 \lesssim \frac{L\tilde\mu^2}{\lambda}\,.
\eeq
One would also expect that medium-induced emissions cannot have a formation time $t_f\sim 2\omega/\vec{k}^2$ much larger than the length of the medium, and thus gluons with
\beq
\omega \gtrsim \frac{\tilde\mu^2L^2}{2\lambda} \equiv \omega_c\,,
\eeq
must have undergone at least one rare hard scattering, with a transverse momentum transfer much larger than the typical value $\tilde \mu$, which is expected to dominate the emission process.\footnote{How one hard scattering dominates the probability of emission can be seen, for example, in figure~2 of \cite{Arnold:2009mr}.} This means that in this high-energy limit the spectrum should be well described by the first order in an opacity expansion (or single hard scattering result), something that was already checked in \cite{Andres:2020vxs}.

On the other hand, gluons with a formation time smaller than the mean free path cannot experience more than one scattering during their formation time. Therefore, the spectrum in the region
\beq
\omega \lesssim \frac{1}{2}\tilde\mu^2\lambda \equiv \omega_{\mathrm{BH}}\,,
\eeq
is also expected to be dominated by single scattering processes. This is the so-called Bethe-Heitler regime \cite{Baier:2000mf}.

The multiple scattering regime is then restricted to gluons emitted with intermediate energies, $\omega_{\mathrm{BH}}<\omega<\omega_c$, where only soft scatterings are expected to occur and approaches such as the harmonic oscillator (HO) approximation and its extensions \cite{Mehtar-Tani:2019tvy,Barata:2020sav} are expected to give a reasonably good description of the in-medium spectrum.

In terms of the parameters from  section~\ref{sec:spectrum}, we can set $\tilde\mu^2 = C\mu^2$ and $\lambda = C/n_0$,  with $C$ a constant factor of order 1.\footnote{The factor $C$ is introduced here in order to match these definitions with the variables to be employed in section~\ref{sec:mult}, where it will be set to $C = e^{-1+2\gamma_E}/4 \approx 0.29$.} Then, the two relevant quantities to calculate the spectrum, $\bar\omega_c$ and $n_0L$, are directly related to the two energy scales described in this section by
\beq\label{eq:omegasyuk}
\bar\omega_c^2=\frac{\omega_{\mathrm{BH}}\,\omega_c}{C^2}\,, \quad \text{and} \quad (n_0L)^2 = C^2\frac{\omega_c}{\omega_{\mathrm{BH}}}\,.
\eeq
In general, the energy spectrum depends only on two dimensionless combinations of $\omega$, $\omega_{\mathrm{BH}}$, and $\omega_c$. We will make use of this property in section~\ref{sec:mult}, where we first define directly $\omega_c$ and $\omega_{\mathrm{BH}}$ in terms of the relevant parameters for the HO approach and later provide the correspondence to the parameters used in the fully resummed evaluation.

\section{Asymptotic limits: single scattering regimes}
\label{sec:asymptotic}

In order to check that the large and small energy limits of the spectrum are dominated by single scattering processes, it is worthwhile looking in detail to the relative sizes of the first and second terms in an opacity expansion. By replacing iteratively eq.~(\ref{eq:SDK}) into (\ref{eq:enspecrinf}) one can generate arbitrary orders of the opacity expansion. For $N=1,2$ one gets, respectively,
\begin{align}
\left.\omega\frac{dI^{\mathrm{med}}}{d\omega}\right|_{N=1} &= \frac{2\alpha_s C_R}{\omega}n_0 \Re\int_0^L ds \int_0^s  dt\int_{\vec{p}\vec{q}}i\,\frac{\vec{p}\cdot\vec{q}}{\vec{q}^2}\sigma(\vec{q}-\vec{p})e^{-i \frac{p^2}{2\omega}(s-t)}\,, \\
\left.\omega\frac{dI^{\mathrm{med}}}{d\omega}\right|_{N=2} &= \frac{2\alpha_s C_R}{\omega} n_0  \Re\int_0^L ds \int_0^s dt\int_{\vec{p}\vec{q}\vec{l}}i\,\frac{\vec{p}\cdot\vec{q}}{\vec{q}^2}\sigma(\vec{q}-\vec{l})e^{-i \frac{l^2}{2\omega}(s-t)} \nonumber\\
&\qquad \times\left[-\frac{n_0}{2}\int_t^s ds'\, \sigma(\vec{l}-\vec{p})e^{-i \frac{p^2-l^2}{2\omega}(s'-t)}\right].
\end{align}
Let us take a closer look at the integration over $\vec{p}$ in the $N=2$ term. Using eq.~(\ref{eq:sigmaV}) we get
\beq
\int_{\vec{p}} \vec{p}\,\sigma(\vec{l}-\vec{p})\,e^{-i \frac{p^2-l^2}{2\omega}(s'-t)} = \int_{\vec{p}} V(\vec{l}-\vec{p})\left[\vec{l}-\vec{p}\,e^{-i \frac{p^2-l^2}{2\omega}(s'-t)}\right]\,.\label{eq:intpsmallom}
\eeq
For $\vec{l}-\vec{p}$ sufficiently small, the phase factor on the right hand side is very close to one and the resulting angular integral is zero by symmetry. In the opposite limit, for large $\vec{p}$ the phase factor oscillates rapidly and does not contribute to the integral. For the asymptotic limits we are interested in, one can then consider the effect of the phase as imposing a lower cut-off $M^2$ (to be determined) and approximate the $\vec{p}$-integration by
\beq
\int_{\vec{p}} \vec{p}\,\sigma(\vec{l}-\vec{p})\,e^{-i \frac{p^2-l^2}{2\omega}(s'-t)} \simeq \;\vec{l}\int_{q^2>M^2} V(\vec{q}) \:\equiv\: \vec{l} \,\Sigma(M^2)\,.
\eeq
The $N=2$ term can then be written as
\begin{align}
\left.\omega\frac{dI^{\mathrm{med}}}{d\omega}\right|_{N=2} &\simeq \frac{2\alpha_s C_R}{\omega} n_0 \Re\int_0^L ds \int_0^s dt\int_{\vec{q}\vec{l}}i\,\frac{\vec{l}\cdot\vec{q}}{\vec{q}^2}\sigma(\vec{q}-\vec{l})e^{-i \frac{l^2}{2\omega}(s-t)} \nonumber \\
&\qquad\times\left[-\frac{n_0}{2}\int_t^s ds'\,\Sigma(M^2)\right]\label{eq:N2}
\,.
\end{align}

For large $\omega$, the phase factor in \eqref{eq:intpsmallom} is close to one unless $\vec{p}$ is also very large, thus setting $M^2\sim 2\omega/L$. $\Sigma$ is then sensitive only to the high-momentum tail of the collision rate $V$, which must be $\sim 1/\vec{q}^4$ for large $\vec{q}$ for any model with point-like scatterings in the UV-limit. Hence $\Sigma\sim\mu^2L/\omega$, showing that the $N=2$ term is suppressed by an additional power of $\omega$ with respect to the $N=1$ term, without any large logarithms to compensate in the large-$\omega$ region. It is easy to see that a similar argument shows that further terms in the opacity expansion are also suppressed by higher powers of $\omega$, meaning that the asymptotically large-$\omega$ limit of the full spectrum is indeed the $N=1$ (also known as GLV or single hard scattering) result.

In the opposite limit, for $\omega$ asymptotically small, the phase factor in \eqref{eq:intpsmallom}, as well as similar factors appearing in higher orders in opacity, oscillates rapidly and does not contribute to the integral except for very small values of $(\vec{p}-\vec{l})^2$. It thus has the effect of cutting  a possible divergence at zero momentum transfer. The scale for this IR cutoff is then set by $M^2\sim l^2$, given that the phase in the first line of \eqref{eq:N2} must be kept. In the absence of an IR divergence in the interaction this cutoff is not necessary, in which case $M^2$ can be safely taken to zero, and thus $\Sigma$ becomes the total collision cross section.

We can interpret the results above in terms of real-virtual cancellations by recalling that the two terms of eq.~(\ref{eq:sigmaV}) correspond, respectively, to the real and virtual contributions for the scattering process. In the case of large $\omega$ the cancellation is almost complete, confirming that the probability of emission for large energies is dominated by single hard scatterings. This does not mean that the gluon is not allowed to have more scatterings, but those subsequent scatterings have the same effect as momentum broadening, which does not affect the probability of emission. On the other hand, for small $\omega$ the formation time picture which restricts the number of scatterings seems to apply only to the real contributions, limiting the possibility for real scatterings only to very small transverse momentum transfers, while the virtual contributions are unconstrained. As usual, the virtual contribution exponentiates, thus resulting in a factor which can be interpreted as the probability of not having further scatterings.

The resummed spectrum at low energies then takes the form
\beq
\left.\omega\frac{dI^{\mathrm{med}}}{d\omega}\right|_{\omega\to 0} = \frac{2\alpha_s C_R}{\omega} \Re\int_0^L ds\; n_0 \int_0^s dt\int_{\vec{p}\vec{q}}i\,\frac{\vec{p}\cdot\vec{q}}{\vec{q}^2}\sigma(\vec{q}-\vec{p})e^{-\left(i \frac{p^2}{2\omega}+\frac{1}{2}n_0\Sigma(p^2)\right)(s-t)}\,,
\label{eq:speclowomega}
\eeq
which clearly has the structure of the $N=1$ result times a no-scattering probability factor.

These suppression factors were already present in sections~5~and~6 of \cite{Wiedemann:2000za}. Although that paper is focused on calculating arbitrary orders in opacity, it can be shown that their ``incoherent limit'', where $L\to\infty$ with $n_0L$ fixed, can be resummed yielding a similar result to our small-$\omega$ limit. The equivalence of both limiting cases can be easily seen from the limit $x\to 0$ with $x$ defined in \eqref{eq:our_x}.

We shift the focus now to performing a numerical comparison between our low-$\omega$ result given by  eq.~\eqref{eq:speclowomega} and the all-order spectrum of section~\ref{sec:spectrum}. The additional factor in eq.~\eqref{eq:speclowomega} with respect to the $N=1$ term does not add any additional difficulty to its numerical evaluation. One can then perform all the integrations analytically except for one, as explained in detail in appendix~\ref{sec:integrals}.

\begin{figure}
\vspace{-5mm}
\centering
\includegraphics[scale=0.40]{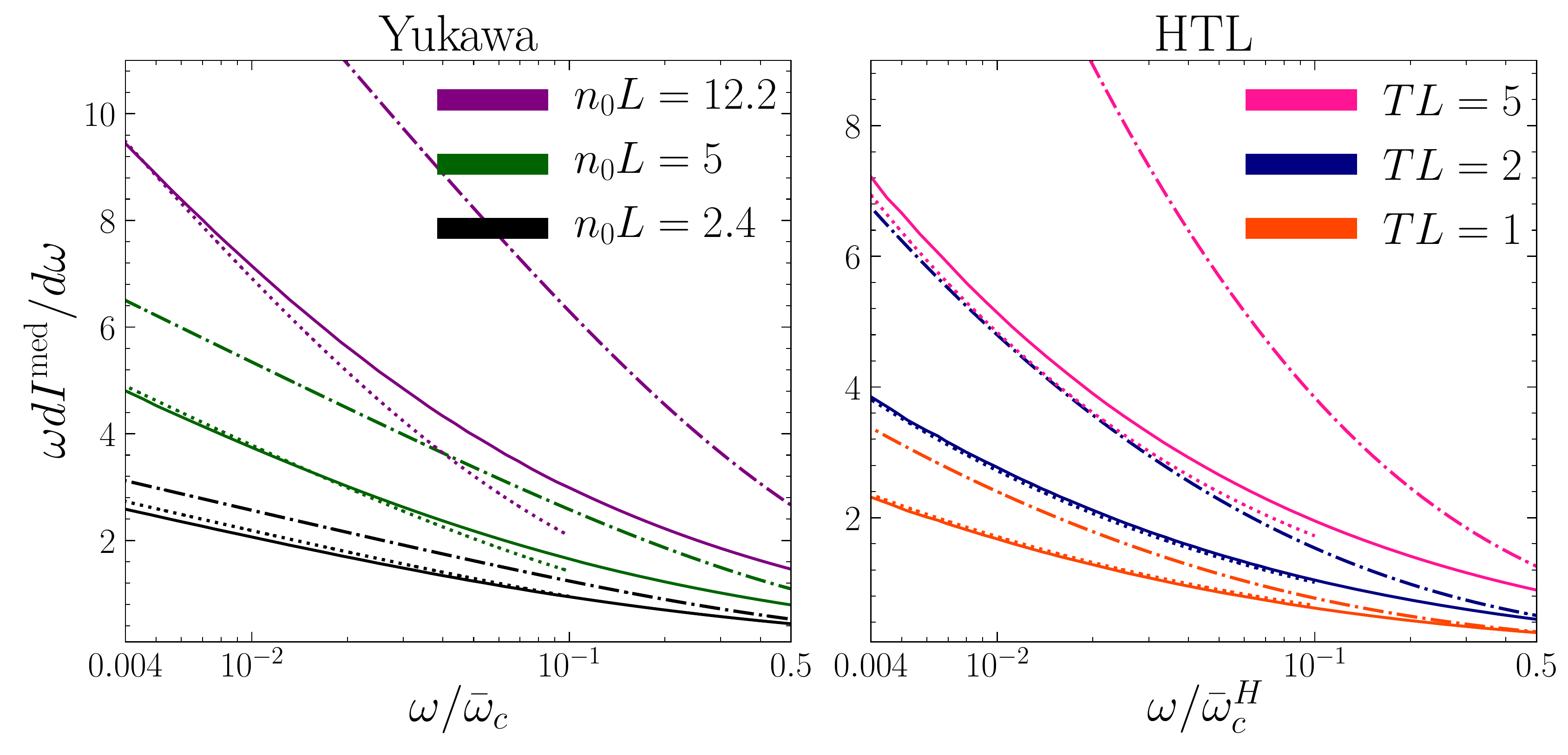}
\caption{Left panel: Fully resummed medium-induced gluon energy distribution for the Yukawa collision rate (solid lines), its low energy limit given by eq.~(\ref{eq:speclowomega}) (dotted), and the GLV $N=1$ result (dash-dotted) as a function of $x=\omega/\bar{\omega}_c=2\omega/(\mu^2L)$ for different values of $n_0L$. Right panel: Fully resummed medium-induced gluon energy distribution for the HTL collision rate (solid lines), its low energy limit given by eq.~(\ref{eq:speclowomega}) (dotted), and the GLV $N=1$ result (dash-dotted) as a function of $x=\omega/\bar{\omega}_c^H=2\omega/(m_D^2L)$ for different values of $TL$.} 
\label{fig:low_energy}
\end{figure}

We also have to specify the collision rate $V$. Similarly to what was presented in \cite{Andres:2020vxs} we consider two different models, a Yukawa-type interaction (also known as Gyulassy-Wang model \cite{Wang:1991xy}) and a hard thermal loop (HTL) interaction \cite{Aurenche:2002pd}, which have radically different IR-behaviors. On one hand, the Yukawa interaction is taken as
\beq
V_Y(\vec{q}) = \frac{8\pi\mu^2}{(\vec{q}^2+\mu^2)^2}\,\label{eq:Vyuk}
\eeq
which is finite for $\vec{q}=0$. So, it yields
\beq
\Sigma_Y(M^2) = \frac{2\mu^2}{M^2+\mu^2}\,,\label{eq:SigmaY}
\eeq
where $M^2$ can be safely taken to 0 and use $\Sigma = 2$ in \eqref{eq:speclowomega} for the asymptotic case.

On the other hand, the HTL interaction is given by
\beq
\frac{1}{2}n_0\,V_H(\vec{q}) = \frac{g_s^2N_c m_D^2T}{\vec{q}^2(\vec{q}^2+m_D^2)}\, ,\label{eq:VHTL}
\eeq
which has an IR-divergence. Thus, it yields
\beq
\frac{1}{2}n_0\,\Sigma_H(M^2) = \alpha_sN_cT\ln\left(1+\frac{m_D^2}{M^2}\right)\,,\label{eq:SigmaH}
\eeq
where it is now clear that the cutoff $M^2$ becomes important.

For illustration purposes, we  assume throughout this manuscript  the parent parton to be a quark, i.e., we take $C_R=C_F=4/3$. The strong coupling is fixed to $\alpha_s=0.3$.

In figure~\ref{fig:low_energy} we present a comparison between the fully resummed evaluation (solid lines), its low-energy limit given by eq.~\eqref{eq:speclowomega} (dotted lines) and the $N=1$ result (dash-dotted lines) for the Yukawa collision rate for three different values of $n_0L$ (left panel), and for the HTL  collision rate for three values of $TL$ (right panel). We have selected the values for $n_0L$ for the Yukawa case and those of  $TL$ for the HTL to coincide when following the correspondence between the two interaction models given in \cite{Andres:2020vxs}. In all cases it is clear that the asymptotic formula derived in this section is very close to the all-order result for sufficiently low energy.

Figure~\ref{fig:low_energy} also shows that even though the precise form of the curves in this low-energy region depends on the details of the parton-medium interaction, the physical picture described in this section works equally well for collision rates either with or without an IR divergence. Our results confirm that while the emission at low energies seems to be dominated by a single real scattering, it is not possible to get the correct result without accounting for the resummation of the virtual contributions, as it can be seen in the great departure of the $N=1$ result with respect to the full evaluation.\footnote{It is important to notice that this regime is at the boundary of the region where the assumptions used in the derivation of eq.~\eqref{eq:bdmps} are expected to hold. Some effects that might be important in this region and have not been taken into account are: exact kinematics (energy not being equal to the longitudinal momentum), thermal masses in the propagators, and multiple emissions.}

\section{Multiple scattering regime}
\label{sec:mult}

We now leave aside the low- and high-energy kinematic regions, where the emission process is dominated by a single scattering, to turn our attention to the intermediate energy region ($\omega_{\mathrm{BH}} < \omega < \omega_c$). In this domain the radiation process is presumed to be mainly controlled by multiple scatterings which are typically resummed through the well known Harmonic Oscillator approximation. 

\subsection{Energy spectrum}
\label{subsec:spectrum}

The HO or multiple soft scattering approach relies on approximating the dipole cross section in coordinate space $v(\vec{r})$ by
\beq
v(\vec{r}) \equiv \frac{n_0}{2}\int_{\vec{q}}
e^{-i\vec{q}\vec{r}}\, \sigma({\vec{q}})
\approx \frac{1}{2} \hat{q}\,\vec{r}^2 \,,
\label{eq:vr}
\eeq
where $\hat{q}$ is the so-called jet quenching parameter that describes the average transverse momentum squared transferred from the medium to the parent parton per unit path length. This approximation gives rise to a simple analytic expression for the in-medium emission spectrum.

In spite of its limitations, the HO spectrum is expected to capture the overall behavior of the gluon emission process at intermediate energies for dense media. Therefore, it is interesting to establish in a rigorous way in which kinematic regime this result provides a proper description of the radiation process. Nevertheless, comparing  the fully resummed and HO spectra is not straightforward since the correspondence between the set of parameters used in each evaluation is not, in principle, unequivocal.

One can show that $\hat{q}$ is related to the first moment of the dipole cross section $\sigma(\vec{q})$. However, when performing this integration for any collision rate with Coulomb-like interactions at short distances one encounters a logarithmic divergence which must be regulated by a cut-off. Thus, the relation between $\hat{q}$ and the parameters entering the collision rate $V(\vec q)$ depends on this cut-off. For phenomenological purposes, since $\hat{q}$ is regarded as a local property of the medium (independent of both the emitter and the radiated gluon), this cut-off is  usually taken as a constant. This choice reduces by one the number of free parameters in the HO approximation with respect to the full evaluation, allowing only a qualitative comparison between both approaches, as done in \cite{Andres:2020vxs}. In this manuscript, we attempt a more quantitative comparison between both evaluations by following the prescription obtained in \cite{Mehtar-Tani:2019tvy}, where the cut-off depends on the energy of the emitted gluon. More specifically, the dipole cross section is set to its leading logarithmic approximation
\beq
v(\vec{r})=
\frac{\hat{q}_0}{4}\vec{r}^2
\ln \left( \frac{1}{\vec{r}^2\mu^{\star2}}
\right)\,,\label{eq:vrlog}
\eeq
which is the correct limit for small dipole separation $\vec{r}$ provided that the collision rate decays (at short distances) as $V(\vec{q}) \sim 1/\vec{q}^4$. Then, $v\left(\vec{r}\right)$ can be decomposed as 
\beq
v(\vec{r}) = \frac{\hat{q}_0}{4}\vec{r}^2 
\left(
\ln \left( \frac{Q^2}{\mu^{\star2} } \right) 
- \ln ( \vec{r}^2Q^2 ) \right)
= v_{\mathrm{HO}}( \vec{r} ) + 
v_{\mathrm{pert}}( \vec{r}) \,,
\eeq
where $Q^2$ is an arbitrary matching scale which plays the role of the aforementioned cut-off, $v_{\mathrm{HO}}\left(\vec{r}\right)$ is the HO approximation of the dipole cross section, and $v_{\mathrm{pert}}\left( \vec{r} \right)$ is treated as a perturbation. It was argued in \cite{Mehtar-Tani:2019tvy} that  $Q^2$ should be taken to be of the order of $Q_c^2$ defined through the following transcendental equation
\beq
Q_{c}^{2}= \sqrt{2\omega \hat{q_0} \ln 
\left(\frac{Q^2_c}{\mu^{\star 2}}\right)} 
\,.
\label{eq:Q_c2}
\eeq
Thus, we will take $Q^2= a\,Q_c^2$, $a$ being  a constant of order $1$.

When the in-medium energy spectrum is computed to the lowest order, taking $v(\vec{r})= v_{\mathrm{HO}}(\vec{r})$,  the well known HO result is recovered
\beq
\omega \frac{d I^{\mathrm{HO}}}{d \omega} =  
\frac {2\alpha_sC_R}{\pi}\,
\ln \left|\cos \left(\Omega L \right) \right| \,,
\label{eq:spectrum_initial_LO}   
\eeq  
where $\Omega$ is given by
\begin{equation}
\Omega=\frac{1-i}{2} \sqrt{\frac{\hat{q}}{\omega}}\,,
\quad  
\mathrm{and}
\quad
\hat{q}=\hat q_0 \ln 
\left(
\frac{Q^2}{\mu^{\star 2}}
\right)
\,.
\label{eq:Omegaqhat}
\eeq

Even with the prescription for the cut-off given in eq.~(\ref{eq:Q_c2}), the HO spectrum is still dependent on the choice of the factor $a$. One solution to this issue is to compute the next order in $v_{\mathrm{pert}}(\vec{r})$, which was also shown to fix the   high-$\omega$ behavior of the spectrum by making it  coincide with the single hard scattering result \cite{Mehtar-Tani:2019tvy}. This next-to-leading order (NLO) contribution is as follows \cite{Barata:2020sav}
\beq
 \omega \frac{d I^{\mathrm{NLO}}}{d \omega}=
 \frac{\alpha_s C_R}{2\pi} \,\hat{q}_0  
 \Re \left\{
 \int_0^L d s \frac{-1}{k^2\left(s\right)}
 \left[ \ln \left(-\frac{k^2\left(s\right)}{Q^2}\right)+\gamma_E
 \right]
\right\} \,,
\label{eq:spectrum_initial_NLO} 
\eeq
where $k^2(s)$ is defined as
\beq
k^{2}(s) = \frac{i\omega}{2}\Omega \left[\cot \Omega s  - \tan \Omega(L-s) \right] \,, \label{eq:k}
\eeq
and $\gamma_E$ is the Euler-Mascheroni constant. Note that the NLO term should be taken from \cite{Barata:2020sav} instead of \cite{Mehtar-Tani:2019tvy}, where its expression contains an incorrect sign.

The in-medium energy spectrum depends on three parameters: $\mu^\star$, $\hat{q}_0$, and $L$. Following the discussion of section~\ref{sec:scales}, we define
\beq
 \omega_c = \frac{\hat q_0L^2}{2}\,,
 \quad \mathrm{and} \quad
 \omega_{\mathrm{BH}} = \frac{\mu^{\star 4}}{2\hat q_0}\,,
\eeq
allowing us to write the spectrum in terms of just the two following  variables
\beq
\chi^2 \equiv \frac{ \omega_c}{ \omega_{\mathrm{BH}}} = 
\left(
\frac{\hat q_0L}{\mu^{\star 2}}
\right)^2\,,
\quad \mathrm{and} \quad
\hat x \equiv \frac{\omega}{\omega_{\mathrm{BH}}}\,.
\label{eq:chi_and_x}
\eeq
In the definition above it is clear that $\chi$ gives a measure of how big is the region where the multiple scattering approach is applicable. It also serves as an estimate of the number of scattering centers encountered by a probe, and thus the multiple soft scattering approximation is expected to work better for larger values of $\chi$.

In terms of these variables, the HO spectrum yields
\beq
\hat x \frac{d I^{\mathrm{HO}}}{d\hat x}=
\frac{2\alpha_s C_R}{\pi}\,
\ln \left|\cos \left(
(1-i)\chi \sqrt{\frac{\Lambda + \ln a}{2 \hat x}}
\right) \right| \,,
\label{eq:LO_spectrum_final}
\eeq
where $\Lambda$ is defined through the following transcendental equation
\beq
\Lambda  = \frac{1}{2} \ln \left( \Lambda \hat x\right)\,.
\label{eq:Lambda}
\eeq
This equation,  as well as (\ref{eq:Q_c2}),  has real solutions only for $\hat x > 2e$, setting the minimum energy for which this formalism holds, and can be solved iteratively.

\begin{figure}
\vspace{-5mm}
\centering
\includegraphics[width=\textwidth]{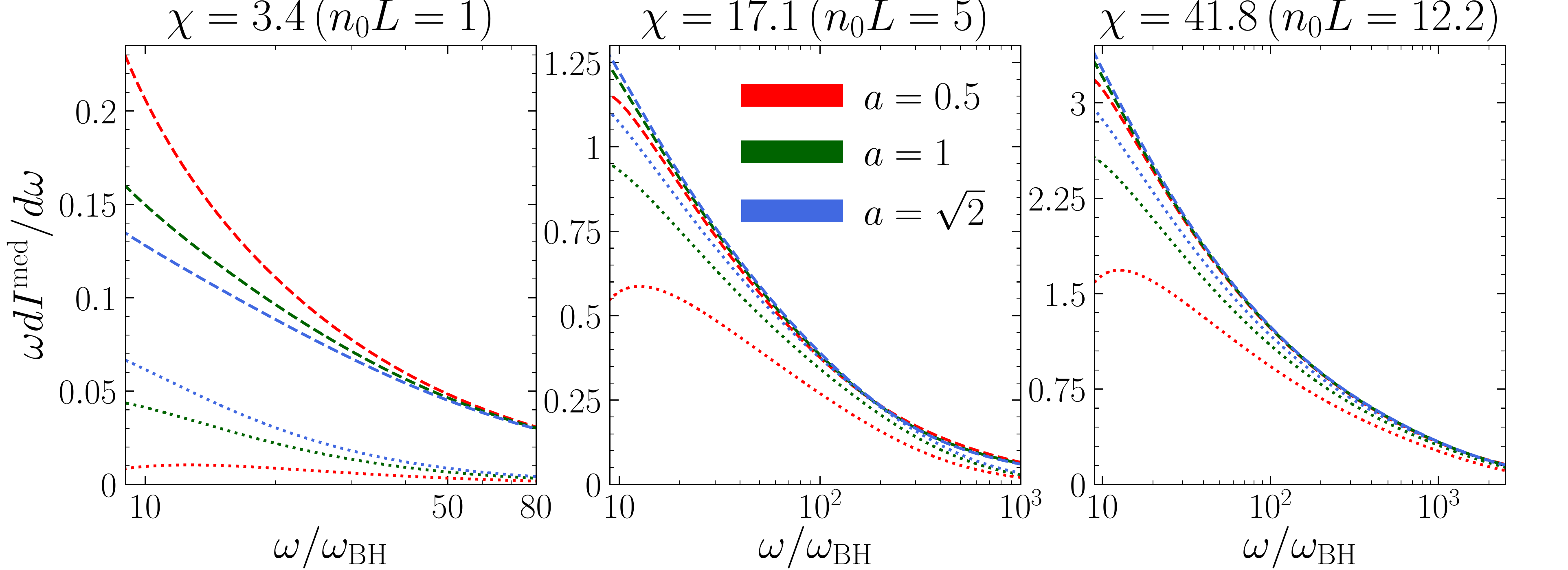}
\caption{Left panel: HO (dotted lines) and HO+NLO (dashed lines) contributions to the gluon energy distribution for $\chi = 3.4$ (or $n_0L=1$) as a function of $\hat{x}=\omega/{\omega}_{\mathrm{BH}} $ for different values of the matching scale $Q^2 =a \,Q_c^2$. Central and right panels: same as left panel for $\chi= 17.1$ (or $n_0L=5$) and $\chi= 41.8$ (or $n_0L=12.2$), respectively.} 
\label{fig:Q2c}
\end{figure}
  
The NLO contribution can be then written as
\beq
\hat x \frac{d I^{\mathrm{NLO}}}{d \hat x}=  \frac{\sqrt{2}\alpha_s C_R}{\pi} 
\Re \left\{
\frac{(i-1)\chi} {\sqrt{\hat x\,(\Lambda +\ln a)}}\int_{0}^{1}  \frac{d s}{K^2(s)} 
\left[\ln \left(-\frac{\left(1+i\right)}{4a\sqrt{2}}\sqrt{\frac{\Lambda + \ln a}{\Lambda}}K^2(s)  \right) +\gamma_{E}\right]
\right\}\,,
\label{eq:NLO_spectrum_final}
\eeq
where $K^2(s)$ is given by
\beq
K^2(s) = \cot \left(
(1-i) \chi\sqrt{\frac{\Lambda+\ln a}{2\hat x}} \,s \right) - \tan \left(
(1-i)\chi\sqrt{\frac{\Lambda +\ln a}{2\hat x}}\,(1-s)
\right)\,,
\label{eq:our_K}
\eeq

Note that since $\Lambda$ depends only on $\hat x$, both the HO and NLO contributions are a function of $\hat x$ and $\chi$ only, together with the specific choice of the value of the constant $a$ entering the matching scale $Q^2$. However, it was argued in \cite{Barata:2020sav} that the dependence on $Q^2$ of the sum of both terms is much weaker. In order to check this claim, we plot in figure~\ref{fig:Q2c} the HO contribution to the spectrum (dotted lines) and the sum of the two lowest orders HO+NLO (dashed lines) as a function of $\hat{x}$ for three different values of  $a$
and for  $\chi=3.4$ (left panel), $\chi=17.1$ (center panel) and $\chi=41.8$ (right panel). While the HO results vary significantly when varying the value of $a$ for all values of $\chi$, it can be seen that the sum of the first two orders is indeed much more stable (only) for large enough values of $\chi$. We have found that this is the case for $\chi\sim 14$ or larger. From now on we will only show the sum of these two terms and the value of $a$ will be set to $a=1$, keeping in mind that no conclusions can be drawn for small values of $\chi$.

It is also important to emphasize that even though eqs.~\eqref{eq:Q_c2}~and~\eqref{eq:Lambda} can be solved for $\hat x>2e$, $Q_c^2$ was originally introduced as a UV-cutoff and therefore cannot be too small, thus further restricting the minimum value of $\hat x$. In practice, results cannot be trusted for $\hat x<10$ and in consequence we will always use this lower limit for the numerical evaluations.

Finally,  we should mention that higher orders of this expansion have already been computed \cite{Barata:2020sav} but they involve much more complex expressions, thus defeating the purpose of having simple analytical formulas which can be easily evaluated. The NNLO correction is claimed to be small with respect to the HO+NLO result and, in consequence, this HO+NLO spectrum should be a good approximation of the fully resummed one. We can check this claim by directly comparing to our all-order result, and thus we do not need to take into account further orders in this expansion.

In order to make a meaningful comparison between the HO+NLO result and the all-order evaluation described in section~\ref{sec:spectrum} and ref.~\cite{Andres:2020vxs}, it is necessary to establish a correspondence between their different sets of parameters. For simplicity, we consider the fully resummed spectrum with the Yukawa-type interaction only, given that for the regime considered in this section the energy spectra obtained with both the Yukawa and HTL interaction models are very similar. This is illustrated in figure~\ref{fig:potentials}, where we have made use of the leading-logarithmic mapping between the parameters entering in both collisions rates derived in \cite{Barata:2020sav}. For further details about this correspondence, we refer the reader to ref.~\cite{Andres:2020vxs}, where this mapping was employed to compare the $\vec{k}$-differential spectra for both parton-interaction models.

\begin{figure}
\vspace{-5mm}
\centering
\includegraphics[scale=0.40]{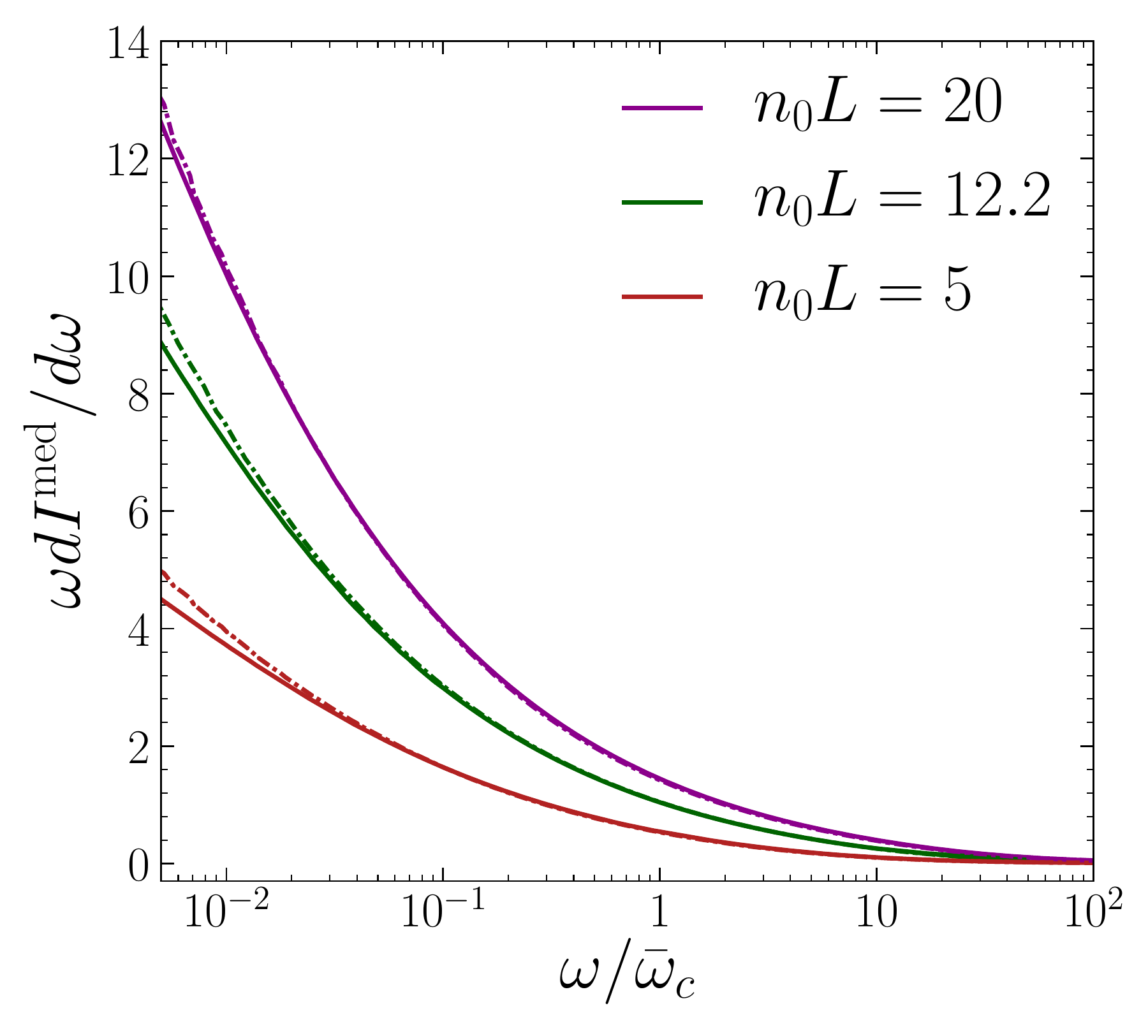}
\caption{Fully resummed medium-induced gluon energy distribution for the Yukawa (solid lines) and  HTL (dash-dotted) collision rates as a function of the rescaled gluon energy $x=\omega/\bar{\omega}_c$ for different values of $n_0L$: $n_0L=20$ (corresponding to $TL=8$), $n_0L=12.2$ (or $TL=5$), and $n_0L=5$ (or $TL=2$).} 
\label{fig:potentials}
\end{figure}

Accordingly, we show how to map the parameters of the all-order evaluation for the Yukawa collision rate (only) with those entering the HO+NLO approximation. As it was explained in \cite{Barata:2020sav}, one can compute  $v(\vec{r})$ exactly, as defined in eq.~\eqref{eq:vr}, using \eqref{eq:sigmaV}~and the Yukawa collision rate (given by \eqref{eq:Vyuk}), and then verify that it indeed has the small-$\vec{r}$ behavior of eq.~\eqref{eq:vrlog} as long as one identifies $\hat q_0 = n_0\mu^2$ and $\mu^{\star 2} = C\mu^2$ with $C = e^{-1+2\gamma_E}/4 \approx 0.29$. Therefore, the correspondence between the actual variables used in the all-order  evaluation, $x$ and $n_0L$, and those appearing in the HO+NLO approximate result, $\hat x$ and $\chi$, is 
\beq
\hat x =  \frac{n_0L}{C^2}\,x\,,  \quad \mathrm{and} \quad
\chi =  \frac{n_0L}{C}\,.
\label{eq:dictionary}
\eeq
This one-to-one mapping enables a quantitative comparison between the HO+NLO and the fully resummed spectrum, which allows us to determine in which regions of the parameter space the HO+NLO expansion is a good approximation of the all-order result.

\begin{figure}
\vspace{-5mm}
\centering
\includegraphics[width=\textwidth]{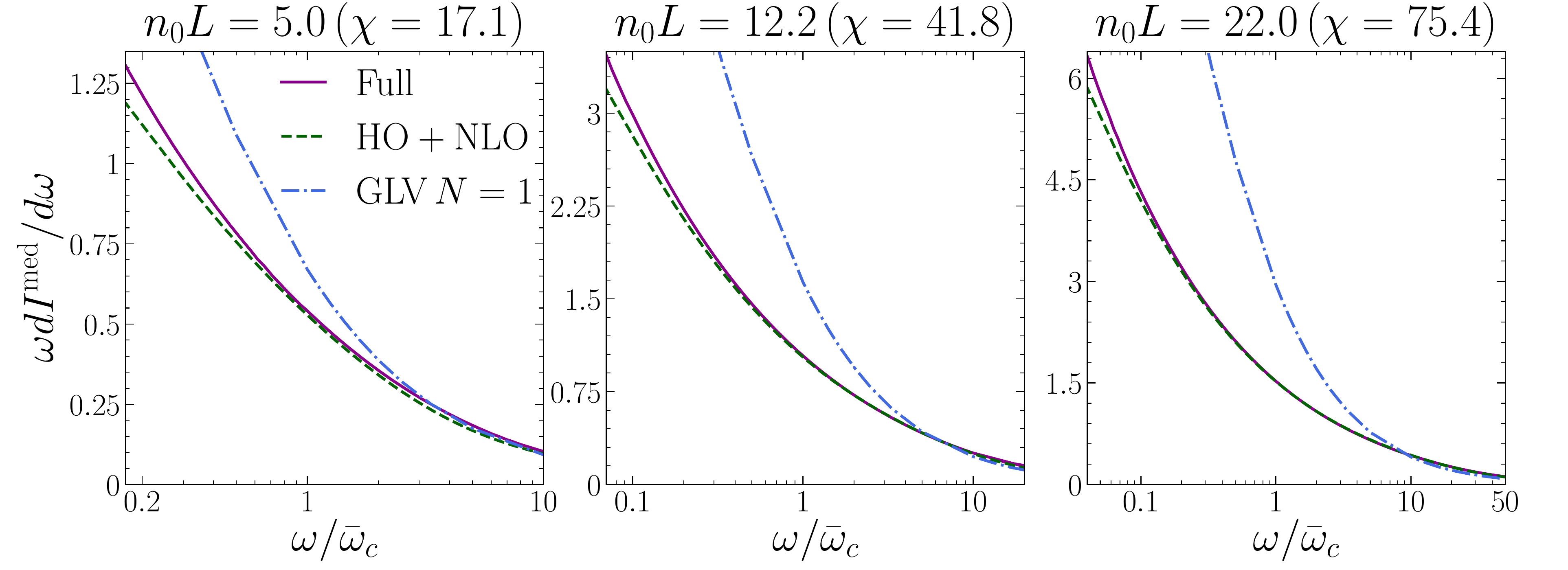}
\caption{Fully resummed medium-induced gluon energy distribution for the Yukawa collision rate (magenta solid lines) compared to the HO+NLO (green dashed) and first opacity (blue dash-dotted) approximations as a function of the rescaled gluon energy $x=\omega/\bar{\omega}_c$ for different values of $n_0L$ (or $\chi$).} 
\label{fig:spectra}
\end{figure}

We present in figure~\ref{fig:spectra} a comparison between our fully resummed energy distribution for the Yukawa collision rate, the first opacity,\footnote{For the formulas used to compute the GLV $N=1$ energy spectrum see for instance  Appendix B of \cite{Salgado:2003gb} or Appendix B of \cite{Andres:2020vxs} (with $\bar R\to\infty$).} and the HO+NLO approximation for three different values of $n_0L$ (or $\chi$). The lower value of $n_0L$ is taken just above the minimum value for which the HO+NLO is stable with respect to the choice of the matching scale, and the higher value of $n_0L$ is taken to coincide with the choice of parameters in \cite{Barata:2020sav}. The limits in the horizontal axes are chosen to show the region where multiple scatterings are believed to be important and the HO+NLO is well defined ($\hat x>10$). In all three panels it can be clearly seen that the HO+NLO result closely resembles the full evaluation, while the single hard scattering approximation does not. Increasing $n_0L$ not only makes the relative differences between the full  and HO+NLO spectra smaller, but also makes the region of applicability of the HO+NLO approach much larger.

Even though we choose to focus on the multiple scattering region, it should be noted that the HO+NLO result also has the correct $1/\omega$-behavior at large energies, while it is well known that the HO approximation alone cannot reproduce these large $\omega$-tails --- this can be seen in more detail in our previous paper \cite{Andres:2020vxs}.

\subsection{Emission rates}
\label{subsec:rates}

Instead of always taking the spectrum as a function of $\hat x$ (or $x$) for fixed $\chi$ (or $n_0L$), we can also use $\chi$ as a variable and fix $\hat x$. This approach is in fact equivalent to studying the dependence of the spectrum on the length of the medium $L$, something that is easier to look at through the emission rates, usually defined as the derivative of the spectrum with respect to $L$.

The instantaneous in-medium emission rate is sometimes interpreted as the probability of a hard parton to radiate a soft gluon at a given time. This interpretation is somehow imprecise since gluons are emitted over an extended formation time and therefore cannot be attributed to a given time coordinate, as explained in \cite{CaronHuot:2010bp}. This is less of a problem for emissions with formation times much smaller than the length of the medium, which is the approximation employed in the AMY formalism \cite{Arnold:2002ja,Jeon:2003gi}, where multiple emissions can be included through rate equations, as done in event generators such as MARTINI \cite{Schenke:2009gb}.

Looking at the emission rates is also very useful for determining when the effect of multiple scatterings becomes important. In the asymptotic case, for an infinitely long medium, the rates should saturate and reach the AMY value calculated in \cite{Jeon:2003gi}, while for short media the single scattering case predicts a linear rise. These features were both confirmed in \cite{CaronHuot:2010bp}.

Computing the in-medium radiation rates in the fully resummed formalism outlined in section~\ref{sec:spectrum} is straightforward. One only has to notice that in eq.~\eqref{eq:enspecrinf} the length of the medium $L$ appears only as the upper limit of the integral over the position of the last scattering center $s$. Thus, taking the derivative with respect to $L$ amounts to not performing this last integration.

\begin{figure}
\centering
\vspace{-5mm}
\includegraphics[width=\textwidth]{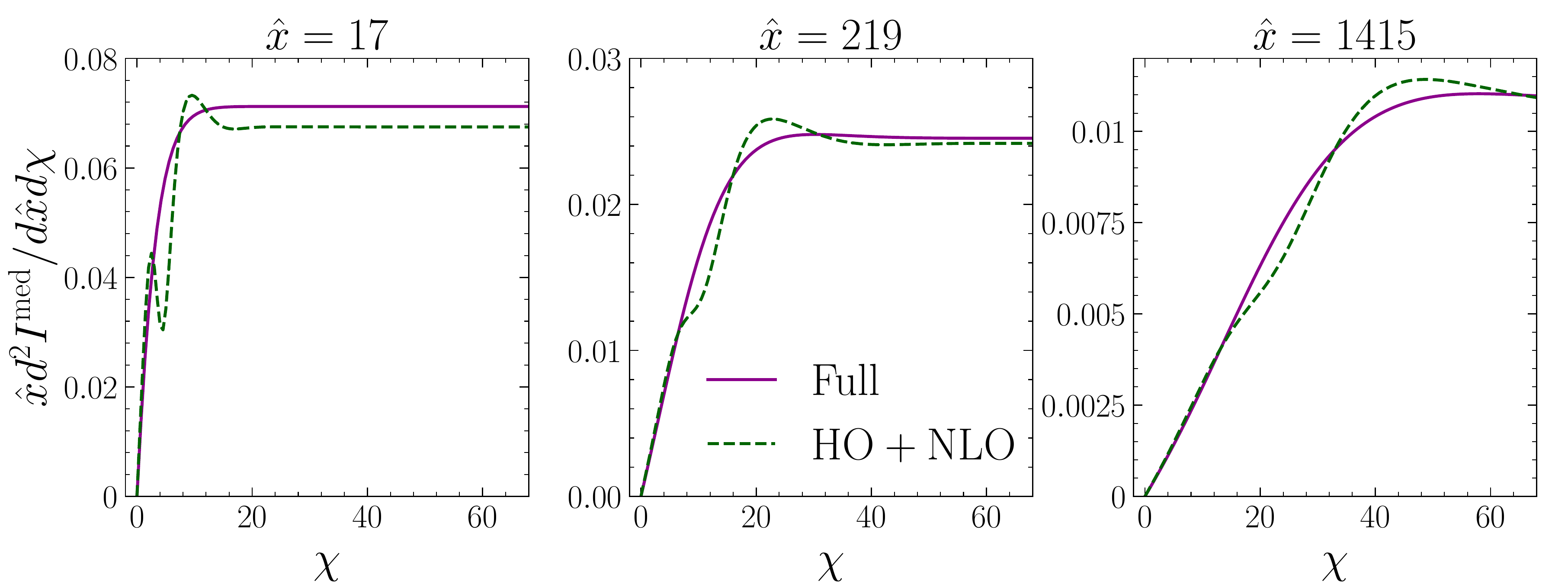}
\caption{Emission rates for the all-order evaluation (magenta solid lines) and the HO+NLO (green dashed) approximation as a function of  $\chi$ for different values of $\hat x =\omega/ \omega_{\mathrm{BH}}$.} 
\label{fig:rates}
\end{figure}

In the case of the analytic expansion outlined in section~\ref{subsec:spectrum} deriving the energy spectrum (before having rescaled its variables) with respect to $L$ is equivalent to performing the derivative with respect to $\chi$ of its rescaled expressions (eqs.~\eqref{eq:LO_spectrum_final}~and~\eqref{eq:NLO_spectrum_final}). For the HO rate we obtain
\beq
\hat x \frac{d^2 I^{\mathrm{HO}}}{d \hat x d \chi} =  \frac{2\alpha_s C_R}{\pi} \Re \left\{(i-1) \sqrt{\frac{\Lambda}{2\hat{x}}}
\tan{ \left((1-i)\chi \sqrt{\frac{\Lambda}{2\hat x}} \right)} 
\right\}\,.
\label{eq:rates_LO_chi}
\eeq
The NLO contribution yields
\begin{multline}
\hat x \frac{d^2 I^{\mathrm{NLO}}}{d \hat x d\chi}=  \frac{\sqrt{2}\alpha_s C_R}{\pi} 
\Re \left\{
\frac{i-1} {\sqrt{\hat x\Lambda}}\int_0^1  \frac{ds}{K^2(s)} \left[\ln \left(-\frac{(1+i)}{4\sqrt 2 }K^2(s)  \right) + \gamma_E \right. \right. \\
\left. \left. + \frac{\chi}{K^2(s)}\frac{dK^2(s)}{d\chi} \left(1 -\ln \left(-\frac{(1+i)}{4\sqrt 2}K^2(s)  \right) - \gamma_E \right) \right]\right\}\,,
\label{eq:rates_NLO_chi}
\end{multline}
where $K^2(s)$ was defined in \eqref{eq:our_K} and its derivative with respect to $\chi$ is given by
\begin{equation}
    \frac{dK^2(s)}{d\chi} = -(1-i)\sqrt{\frac{\Lambda}{2\hat{x}}} \left[\frac{s}{\sin^2{\left((1-i)\chi\sqrt{\frac{\Lambda}{2\hat{x}}}s\right)}} + \frac{1-s}{\cos^2{\left((1-i)\chi\sqrt{\frac{\Lambda}{2\hat{x}}}(1-s)\right)}} \right]\,.
\end{equation}

We present in figure~\ref{fig:rates} a comparison between the all-order in-medium emission rate and the HO+NLO result for three different values of $\hat x$. Both approaches have the expected behavior of first growing linearly for small values of $\chi$ and then reaching a constant asymptotic value for large values of $\chi$, but there are some clear differences. First, the HO+NLO presents some oscillations in the transition region which get worse for lower values of $\hat x$, contributing to the discrepancies we have already seen for lower values of $\chi$ in the spectrum in figure~\ref{fig:spectra}. Second, even though both approaches saturate and become constant at large $\chi$, those asymptotic values do not coincide exactly, with the difference being larger again for smaller values of $\hat x$. For larger values of $\hat x$, the asymptotic values seem to be in better agreement but then these are reached only for very large values of $\chi$, as can be seen in the right panel.

The asymptotic limit of the rates can be directly calculated in both approaches. For the fully resummed evaluation, one has to integrate eq.~\eqref{eq:diffeqF} over $t$ from $0$ to $\infty$. In terms of $\hat x$ and $\chi$, the asymptotic in-medium all-order rate can be written as
\beq
\lim_{\chi \rightarrow \infty} \hat x \frac{d^2 I}{d \hat x d \chi} =
4\alpha_s C_R \Re \int_{\vec{p}} i\vec{p} \cdot \,\vec{G}_{\hat x}(p)\,,
\eeq
where $\vec{G}_{\hat x}(\vec{p})$ satisfies
\beq
 \int_{\vec{q}} \frac{\vec{q}}{\vec{q}^2}\,\tilde \sigma(\vec{q}-\vec{p};\hat{x})  = \frac{ip^2}{C} \vec{G}_{\hat x}(\vec{p}) \,
+ \frac{C}{2} \int_{\vec{q}} \tilde \sigma(\vec{p}-\vec{q};\hat x) \,\vec{G}_{\hat x}(\vec{q}) \,,
 \label{eq:integraleq_G}
\eeq
which can be easily solved numerically and coincides with the soft limit of the AMY result (see eqs.~6~and~7 of \cite{Jeon:2003gi}).
\begin{figure}
\vspace{-5mm}
\centering
\includegraphics[scale=0.40]{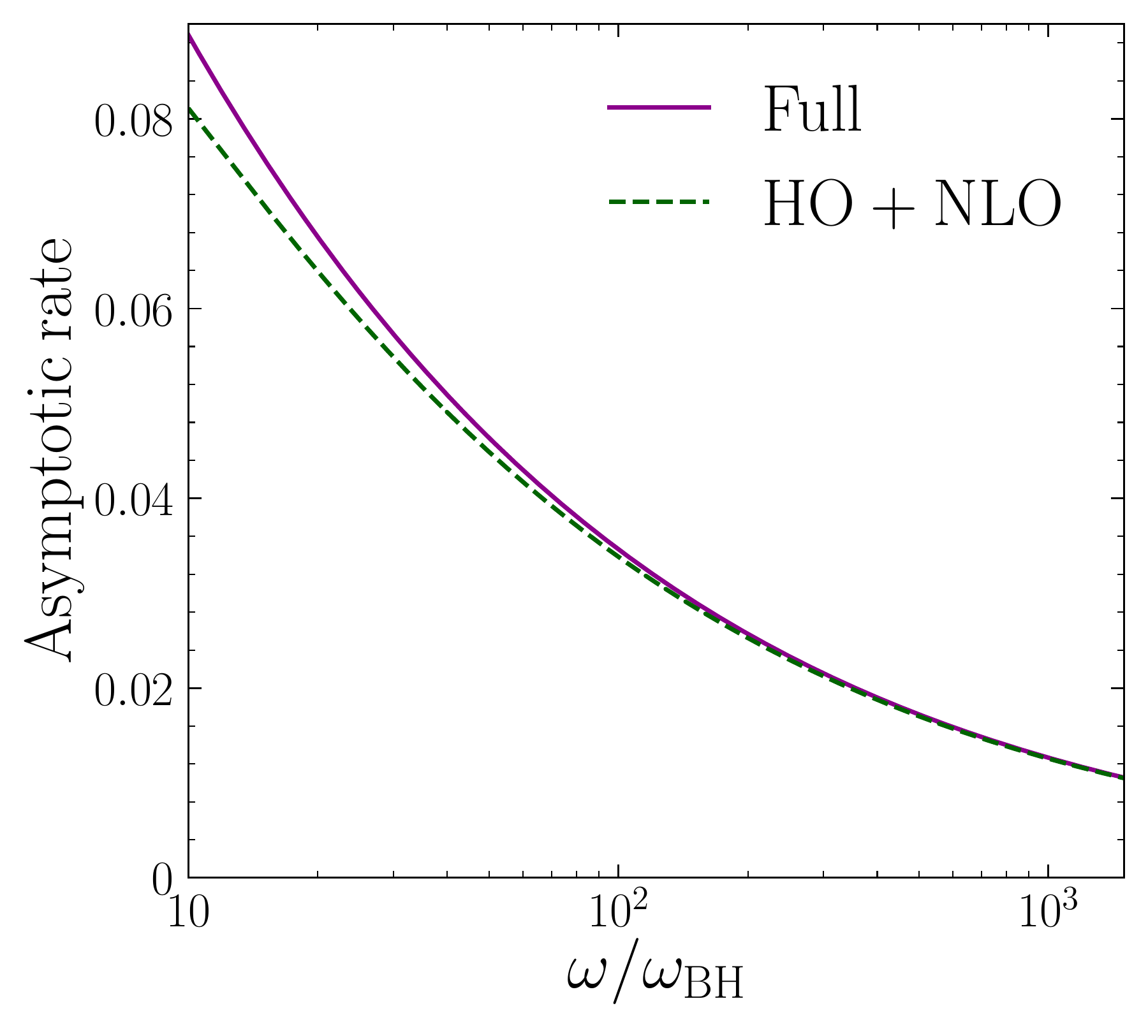}
\caption{Asymptotic rates for the all-order evaluation (magenta solid lines) and the HO+NLO (green dashed) approximation as a function of  $\hat x =\omega/ \omega_{\mathrm{BH}}$.} 
\label{fig:asym}
\end{figure}

For the analytic expansion \cite{Mehtar-Tani:2019tvy}, the asymptotic values of the rates are obtained by taking the $\chi \rightarrow \infty $ limit of eqs.~(\ref{eq:rates_LO_chi})~and~(\ref{eq:rates_NLO_chi}) yielding, respectively,
\beq
\lim_{\chi \rightarrow \infty} \hat x \frac{d^2 I^{\mathrm{HO}}}{d \hat x d \chi}=
\frac{\alpha_s C_R}{\pi} \sqrt{\frac{2\Lambda}{\hat x}}\,,
\label{eq:rateLO_asym}
\eeq
and
\beq
\lim_{\chi\rightarrow \infty} \hat x \frac{d^2 I^{\mathrm{NLO}}}{d \hat x d \chi} = \frac{\alpha_s C_R}{\pi} \frac{ 1}{{\sqrt{2\Lambda \hat x}}}\left[\gamma_E - \ln 2  + \frac{\pi}{4} \right]\,.
\label{eq:rateNLO_asym}
\eeq

We show in figure~\ref{fig:asym} the asymptotic value of the HO+NLO and  all-order in-medium radiation rates as a function of $\hat x$. As observed before, the HO+NLO result is close to the full one specially for large gluon energies ($\hat x > 200$). This large-energy region corresponds to the center and right panels of figure~\ref{fig:rates}, where it is clear that the rates only reach their asymptotic values for very large vales of $\chi$ (or, equivalently, for very large values of $n_0L$).

\section{Conclusions}
\label{sec:conclusions}

In this manuscript we have made use of the all-order formalism for medium-induced gluon radiation derived in \cite{Andres:2020vxs} to shed light on the different physical processes dominating across the different energy scales, with special emphasis on determining where  multiple scattering effects are important.

We have obtained the low-energy asymptotic limit of the full-resummed evaluation given by eq.~(\ref{eq:speclowomega}), which matches the full solution for sufficiently low energies ($\omega <  \omega_{\mathrm{BH}}$). Moreover, (\ref{eq:speclowomega}) provides a very clear physical picture of this low-$\omega$ region: the spectrum can be interpreted as a single in-medium scattering times the probability of not having any further scatterings. It is clear then that, even though the process is dominated by one scattering, accounting for multiple scattering effects is crucial to get the correct behavior of the spectrum in this region through the resummation of the virtual terms. Only including the effect of the $N=1$ opacity expansion gives a huge overestimation of the spectrum.

\begin{figure}
\vspace{-5mm}
\centering
\includegraphics[scale=0.40]{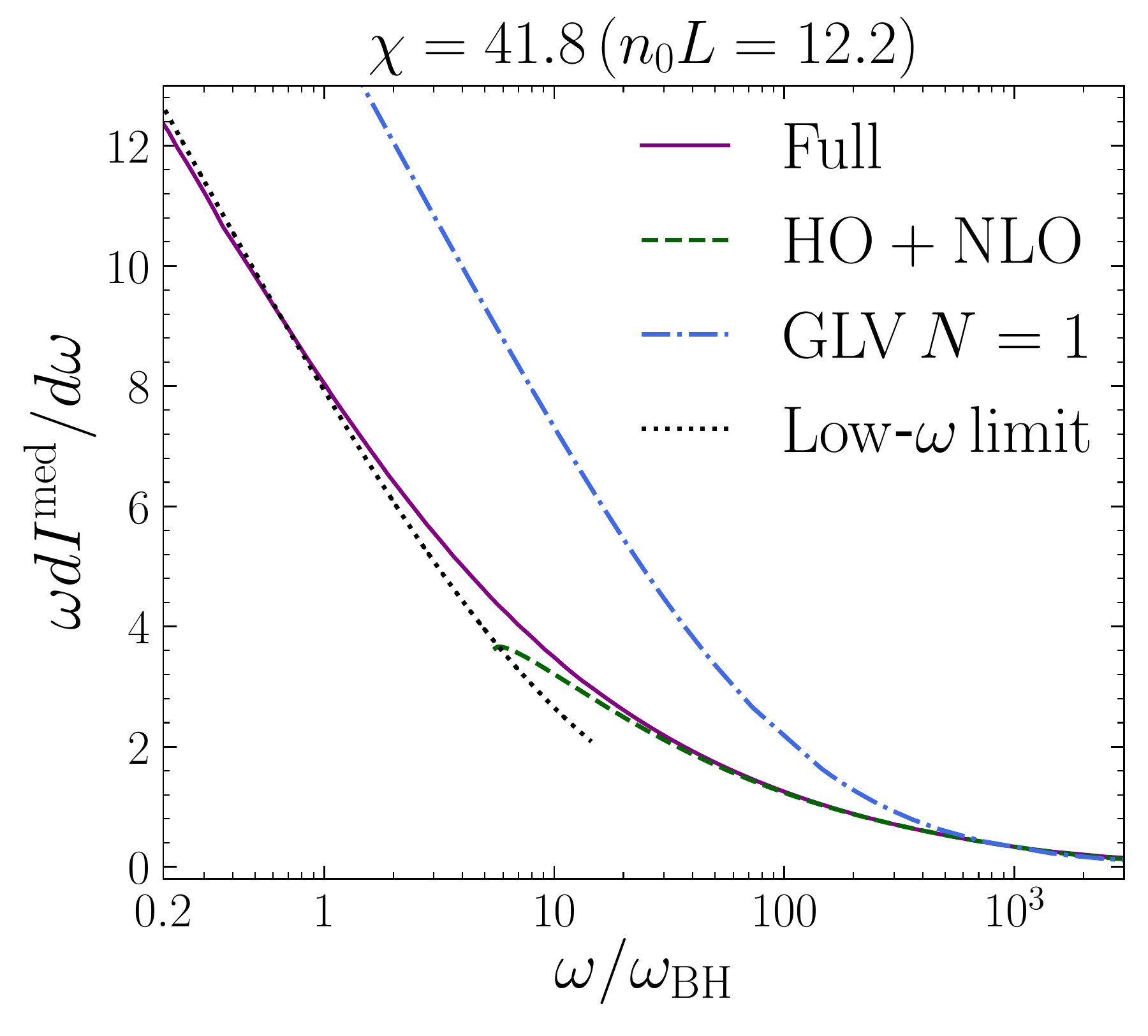}
\caption{All-order medium-induced gluon energy spectrum  (magenta solid line) compared to the HO+NLO approximation (green dashed), GLV $N=1$  result (blue dash-dotted), and the low energy limit of the full resummation described in section~\ref{sec:asymptotic} (black dotted) as a function of $\hat x=\omega/ \omega_{\mathrm{BH}}$.} 
\label{fig:conclusions}
\end{figure}

In the intermediate energy region, where the dynamics of the emission process is expected to be dominated by multiple soft scatterings, the expansion derived in \cite{Mehtar-Tani:2019tvy} provides us with a tool to perform a quantitative comparison between our all-order formalism and an analytic approximation including multiple scatterings. The usual approach for accounting for the effect of multiple soft scatterings in this regime is the HO approximation, but comparisons between the all-order evaluation and the HO had proved difficult given the ambiguity in the momentum scale entering the definition of $\hat q$. Nevertheless, the expansion suggested in \cite{Mehtar-Tani:2019tvy} reduces drastically the dependence on the matching scale and enables a direct correspondence between the parameters involved in the different evaluations. The HO+NLO in-medium energy spectrum gives a good description of the full result on its range of applicability ($\omega > 10\,  \omega_{\mathrm{BH}}$ and $\chi > 14$) and is a much better approximation of the full evaluation than the GLV $N=1$ result in this region, thus highlighting the importance of coherence effects among multiple scatterings. We also analyzed the corresponding emission rates, confirming that both approaches agree in most of the relevant features, while also showing us more clearly where the agreement between approaches starts to fail.

In the high energy region ($\omega > \omega_c$) the radiation process is expected to be dominated by a single hard in-medium scattering. In this asymptotic limit the all-order spectrum indeed coincides with the first opacity result, as was already checked in \cite{Andres:2020vxs}, and, furthermore, the HO+NLO spectrum has also the correct high-$\omega$ behavior, which cannot be reproduced by the HO approximation only.

The summary of our results is illustrated in figure~\ref{fig:conclusions}. We can also see in this figure that there is a region, corresponding to energies between $\omega \sim 3\,  \omega_{\mathrm{BH}}$ and $\omega \sim 10\,  \omega_{\mathrm{BH}}$, where none of the approximate results gives a satisfactory description of the full evaluation. This issue seems to be a consequence of the leading logarithmic approximation in which the HO+NLO approach relies and it remains to be seen if further analytical developments in this direction can help to expand the reach of this approximation.

\acknowledgments
We thank Liliana Apolin\'ario, Carlos A. Salgado, Guilherme Milhano, Jo\~ao Barata, N\'estor Armesto, and Al Mueller for useful discussions. This work was supported  by Ministerio de Ciencia e Innovaci\'on of Spain under project FPA2017-83814-P; Unidad de Excelencia Mar\'{\i}a de Maetzu under project MDM-2016-0692; Xunta de Galicia
under project ED431C 2017/07; Conseller\'{\i}a de Educaci\'on, Universidade e Formaci\'on Profesional as Centro de Investigaci\'on do Sistema universitario de Galicia
(ED431G 2019/05); European Research Council under project ERC-2018-ADG-835105 YoctoLHC; and FEDER.
C.A. was supported by FCT Portugal under the project CERN/FIS-PAR/0024/2019. M.G.M. was supported by Ministerio de Universidades of Spain through the National Program FPU (grant number FPU18/01966).

\appendix
\section{Explicit integrations in the small-$\omega$ formula}
\label{sec:integrals}

The explicit evaluation of eq.~\eqref{eq:speclowomega} can be divided in two pieces: the integration over $\vec{q}$, and the time integrations over $s$ and $t$. Let us start with the former,
\beq
S(p^2) = \int_{\vec{q}}\frac{\vec{p}\cdot\vec{q}}{\vec{q}^2}\sigma(\vec{q}-\vec{p}) = \int_{\vec{q}}\left(1-\frac{\vec{p}\cdot\vec{q}}{\vec{q}^2}\right)V(\vec{q}-\vec{p})\,.
\eeq
Taking into account that $V(\vec{q})$ is a function of $\vec{q}^2$ only, $V(\vec{q}) = \hat V(\vec{q}^2)$, one gets
\begin{align}
S(p^2) &= \frac{1}{(2\pi)^2}\int_0^{2\pi}d\theta\int_0^\infty dq\,(q-p\cos\theta)\,\hat V\left(p^2+q^2-2pq\cos\theta\right) \nonumber\\
&= \frac{1}{4\pi}\int_{p^2}^\infty du\,\hat V(u) = \Sigma(p^2)\,,
\end{align}
where $\Sigma$ is given here by its full expression, even for collision rates which are not singular for small momentum transfers, in order to guarantee the convergence of the integral over $p$ for large momenta.

Now, for the time integrations we have
\begin{align}
T(p^2) &= \int_0^L ds\int_0^s dt\, i\,e^{-\left(i\frac{p^2}{2\omega}+\frac{1}{2}n_0\Sigma(p^2)\right)(s-t)} \nonumber\\
&=\frac{L}{\frac{p^2}{2\omega}-\frac{i}{2}n_0\Sigma(p^2)}-\frac{i}{(\frac{p^2}{2\omega}-\frac{i}{2}n_0\Sigma(p^2))^2}\left[e^{-\left(i\frac{p^2}{2\omega}+\frac{1}{2}n_0\Sigma(p^2)\right)L}-1\right]\,.\label{eq:timeintsmallom}
\end{align}

These expressions, combined with the explicit form of $\Sigma$ for each collision rate  given in eqs.~\eqref{eq:SigmaY} and~\eqref{eq:SigmaH}, allow us to evaluate \eqref{eq:speclowomega} where the integration over $p$ must be done numerically. It is worth noting that for asymptotically small values of $\omega$ the second term in \eqref{eq:timeintsmallom} is much smaller than the first one and can therefore be neglected.

\bibliographystyle{JHEP}

\bibliography{references}
\end{document}